\documentclass[useAMS,usenatbib]{mn2e}
\usepackage{latexsym}
\usepackage{times}
\usepackage{subeqn}
\usepackage[german,english,american]{babel}
\newcommand{\apj}{ApJ}
\newcommand{\apjl}{ApJL}
\newcommand{\nat}{Nature}
\newcommand{\aap}{A\&A}
\newcommand{\mnras}{MNRAS}

\def\lesssim{\mathrel{\hbox{\rlap{\hbox{\lower4pt\hbox{$\sim$}}}\hbox{$<$}}}}
\def\gtrsim{\mathrel{\hbox{\rlap{\hbox{\lower4pt\hbox{$\sim$}}}\hbox{$>$}}}}

\title[The non-linear dependence of $F_{\nu}$ on $M$ and $\dot{m}$]{The
Non-Linear Dependence of Flux on Black Hole Mass and Accretion Rate in Core
Dominated Jets}

\author[S.~Heinz and R.~A.~Sunyaev]{S.~Heinz and R.~A.~Sunyaev\\
Max-Planck-Institut f\"{u}r Astrophysik, Postfach 1317, 85741 Garching,
Germany}

\begin{document}
\date{25 July 2003} \maketitle

\begin{abstract}We derive the non-linear relation between the core flux
$F_{\nu}$ of accretion powered jets at a given frequency and the mass $M$
of the central compact object.  For scale invariant jet models, the
mathematical structure of the equations describing the synchrotron emission
from jets enables us to cancel out the model dependent complications of jet
dynamics, retaining only a simple, {\em model independent} algebraic
relation between $F_{\nu}$ and $M$.  This approach allows us to derive the
$F_{\nu}$--$M$ relation for any accretion disk scenario that provides a set
of input boundary conditions for the magnetic field and the relativistic
particle pressure in the jet, such as standard and advection dominated
accretion flow (ADAF) disk solutions.  Surprisingly, the mass dependence of
$F_{\nu}$ is very similar in different accretion scenarios.  For typical
flat--spectrum core dominated radio jets and standard accretion scenarios
we find $F_{\nu}\sim M^{17/12}$.  The 7--9 orders of magnitude difference
in black hole mass between microquasars and AGN jets imply that AGN jets
must be about 3--4 orders of magnitude more radio loud than microquasars,
i.e., the ratio of radio to bolometric luminosity is much smaller in
microquasars than in AGN jets.  Because of the generality of these results,
measurements of this $F_{\nu}-M$ dependence are a powerful probe of jet and
accretion physics.  We show how our analysis can be extended to derive a
similar scaling relation between the accretion rate $\dot{m}$ and $F_{\nu}$
for different accretion disk models.  For radiatively inefficient accretion
modes we find that the flat spectrum emission follows $F_{\nu} \propto
\left(M\,\dot{m}\right)^{17/12}$.
\end{abstract}
\section{Introduction}
Relativistic jets are collimated outflows from the innermost regions of
accretion disks around black holes and neutron stars.  Not all accreting
compact objects form jets, but when they do, the jet synchrotron radiation
typically dominates the radio spectrum of the compact object.  Such objects
are called radio loud.

Compact objects span 9 orders of magnitude in central mass: in active
galactic nuclei (AGN), it ranges from $M\sim10^6 M_{\odot}$ to $M\sim
few\times 10^9 M_{\odot}$, while Galactic X-ray binaries extend this range
down to a few $M_{\odot}$.  Yet, the jets formed by these objects appear
morphologically remarkably similar, and their core emission typically
follows the same flat powerlaw spectrum.  This suggests that the process of
jet formation is universal, and that jets from supermassive black holes of
different masses are not qualitatively different from each other and from
jets in X-ray binaries, called microquasars.  If this is so, we can compare
jets from objects of different mass $M$ (which is measurable dynamically)
and accretion rate $\dot{M}$ (which is proportional to the accretion
luminosity $L_{\rm acc}$) and determine how their observable
characteristics change with $M$ and $\dot{M}$ \citep{sams:96}.

The most readily available observable parameter is the jet flux at a given
frequency, $F_{\nu}$.  In this {\em letter}, we shall therefore derive,
from theoretical arguments, the relationship between $F_{\nu}$ and $M$.  As
we will show, the mathematical structure of the expression for the jet
synchrotron flux $F_{\nu}$ enables us to contract all the model dependent
complications of jet physics into the formula for the observable spectral
index $\alpha$ and thus remove them from the relation between $F_{\nu}$ and
$M$.  For any observed value of $\alpha$ and for a set of boundary
conditions delivered by accretion disk theory, we can thus formulate a {\em
model independent}, non-linear relation between $F_{\nu}$ and $M$.

A lot of effort has gone into searching for observational correlations
between $F_{\nu}$, $M$, and $\dot{M}$.  Such measurements are difficult
because of numerous selection effects.  Nevertheless, some observational
evidence of a non-linear dependence between AGN radio flux and black hole
mass \citep{franceschini:98,laor:00,mclure:01,lacy:01} exists in the recent
literature.  However, other authors have found no such evidence
\citep{ho:02,woo:02}.  A systematic difference of radio loudness between
neutron star and black hole X-ray binaries has also been suggested
\citep{fender:01}.

Because of the large mass difference, any non-linearity between $F_{\nu}$
and $M$ must be most apparent when comparing microquasars with AGN jets.
And indeed, observations show that the radio loudness parameter $R=L_{\rm
R}/L_{\rm UV,X-ray}$, defined as the ratio of radio luminosity (emitted by
the jet) to UV/X-ray luminosity (emitted by the accretion disk) is much
smaller for microquasars during outburst than it is for radio loud AGNs
\citep{falcke:96}.  In other words: the radio jet flux $F_{\nu}$ depends
non-linearly on $M$.

In the following sections, we will argue that this observational
non-linearity is in not only consistent with, but required by the {\em
model independent} $F_{\nu}$--$M$ relation that we will derive below.

\section{Scale invariant jets}
\label{sec:selfsimilar}
The two fundamental parameters that determine the conditions in the inner
accretion disk are the accretion rate $\dot{M}$ and the mass $M$ of the
central object.  All length scales are proportional to the fundamental
scale of the compact object, $r_{\rm g} \propto M$.  The characteristic
accretion rate of the disk is set by the Eddington rate $\dot{M}_{\rm
Edd}$, and for convenience we will define the dimensionless accretion rate
$\dot{m}\equiv\dot{M}/\dot{M}_{\rm Edd}$.  All dynamically important
variables are determined by these two parameters.

Because jets are formed in the inner disk, it is reasonable to assume that
the conditions in the inner jet are set by the conditions in the inner
disk, and thus, they will similarly depend on $M$ and $\dot{m}$.  However,
it is possible that jets are powered by black hole spin extraction
\citep{blandford:77}, in which case all jet variables would also depend on
the black hole spin parameter $a$.  Thus, any dynamically important jet
variable at the base of the jet will be determined by these three
parameters.

\subsection{Dependence on $M$}
As mentioned in the introduction, observations suggest that the process of
jet formation is universal, and that no qualitative difference exists
between jets from objects of different mass.  This morphological and
spectral similarity of jets from objects with fundamentally different black
hole masses suggests that jet formation and propagation might be a scale
invariant processes
\footnote{The scale invariance assumed for the jet structure is only valid
in the inner regions of the jet, where interactions with the environment
are not important.  On large scales, where this interaction dominates the
jet dynamics, additional parameters independent of the inner accretion disk
enter, most prominently the external density and pressure.  In this case,
it is still possible to write down extended scaling relations
\citep{heinz:02b}, but not in the form of eq.~(\ref{eq:self-similar}).
However, since we restrict our analysis to the emission from the jet core,
we can neglect these complications.},
i.e., that there is one relevant length scale in jet formation, which
is $r_{\rm g}$, and that jet dynamics are invariant under changes in
this length scale\footnote{While the influence of the spin parameter
$a$ on jet formation is not clear, it is important to note that a
second length scale might be present in the process of jet formation,
which is the light cylinder radius $r_{\Omega}$ of the black hole
spin. However, $r_{\Omega}$ depends linearly on $M$ and is thus a
multiple of $r_{\rm g}$. The proportionality factor depends {\em only}
on $a$, not on $M$, {\em i.e.}, $r_{\Omega}=f(a)\,r_{\rm
g}$. Thus, for a fixed spin parameter $a$, the only relevant length
scale that changes upon variations in $M$, which we are primarily
concerned with here, is $r_{\rm g}$. This is the length scale of
relevance for changes in $M$, and we will henceforth assume that jet
formation is invariant under changes in $r_{\rm g}$ for otherwise
fixed parameters, as suggested by the observational similarity of jets
from very different $M$. If $a$ does indeed have significant relevance
for the process of jet formation, then variations in $a$ will simply
introduce intrinsic scatter to the relations derived below, as all
relations will be derived for fixed, but arbitrary, $a$.}.

Scale invariance implies that the spatial variation of important jet
quantities (such as the shape of the jet, i.e., its lateral cross section,
the orientation of magnetic field lines, the field strength, etc.) depends
only on the dimensionless variable ${\mathbf r}/r_{\rm g}$.  Thus, a given
variable $f$ should be proportional to a function $\psi({\mathbf r}/r_{\rm
g})$ which depends on ${\mathbf r}$ only through ${\mathbf r}/r_{\rm g}$.
In other words, we can scale a jet model for mass $M_1$ by a length factor
$M_2/M_1$ and some spatially independent normalization factor and arrive at
a jet model for mass $M_2$.

In mathematical terms, this can be expressed as the condition that we
can write any dynamically relevant quantity $f$, such as the magnetic
field $B(r)$, as the product of two decoupled functions:
\begin{eqnarray}
	f(M,\dot{m},a,r) & = & \phi_{f}(M,\dot{m},a)\,
	\psi_{f}\left(\frac{r}{r_{\rm g}},\dot{m},a\right)\nonumber \\ & =
	& \phi_{f}(M,\dot{m},a)\, \psi_{f}(\chi,\dot{m},a)
\label{eq:self-similar}
\end{eqnarray}
where $r$ is the distance to the central engine measured along the jet,
$\phi_{f}$ describes the dependence of $f$ on the central engine mass $M$,
and $\psi_{f}$ describes the spatial dependence of $f$ on the similarity
variable $\chi \equiv r/r_{\rm g}$ for a given set of $\dot{m}$ and $a$.
Note that this is a {\em requirement} we place on the jet model, inspired
by the observational similarity between jets from different kinds of
objects. Not all possible jet model must necessarily satisfy this relation.
However, those models that do satisfy it span an important sub-class of jet
models and all of them will obey the relations derived below.  One
important example of such a model is the \citet{blandford:79} model.

The normalization functions $\phi_f$ reflect the dependence of the
conditions at the base of the jet on the central mass $M$.  Since jets
are launched above accretion disks, it is natural to assume that these
functions $\phi_{f}$ can be adopted from accretion disk models.

For any geometric quantity, such as the jet diameter $R(r)$, the direct
proportionality of $R$ to $r_{\rm g}$ requires that $\phi_{R} \propto M$
(where we have contracted the constant of proportionality into $\psi_{R}$
for convenience).  As a dimensionless variable, it is reasonable to assume
that the jet Lorentz factor $\Gamma$ is entirely
independent\footnote{Comparison with young stellar object jets indicates
that the specific velocity $\Gamma v$ of jets is related to the
orbital/escape speed of the innermost regions of the disk, which is
independent of $M$ for black holes.} of $M$.  While measuring the bulk
velocity of jets directly is impossible in most cases, the existing
observational limits suggest that jets from microquasars are not any more
or less relativistic than their AGN counterparts, despite 9 orders of
magnitude difference in $M$.  We take this as sufficient evidence to assume
in the following that $\Gamma$ does not depend explicitly on mass, which
allows us to write $\Gamma(r)=\psi_{\Gamma}(\chi,\dot{m},a)$, i.e.,
$\phi_{\Gamma}=1$.

The state of the inner accretion disk depends on the accretion rate
$\dot{m}$.  In all standard accretion disk models, the fundamental
quantities take on a rather simple scaling with the black hole mass $M$.
For high accretion rates, where electron scattering becomes the dominant
opacity source and radiation pressure dominates the energetics in the inner
disk, the density and the pressure scale inversely with mass, $\varrho
\propto P_{\rm rad} \propto M^{-1}$ \citep{shakura:76}.  The magnetic field
might be of the same order as the total pressure, and thus $B\propto
M^{-1/2}$ \citep{shakura:73}.  Recent numerical computations of the
magneto--rotational instability in accretion disks seem to support this
statement \citep{balbus:98}.  The same scaling arises in advection
dominated flows \citep{narayan:95}, and in any scenario where dynamical
terms dominate the cooling rate (e.g., convection or outflow dominated
disks).  Thus, we have $\phi_{P}=M^{-1}$, $\phi_{\varrho}=M^{-1}$, and
$\phi_{B}=M^{-1/2}$.  Observations of X-ray binaries suggest that jet
formation is linked to the so called {\em low-hard state}
\citep{fender:01}, which is believed to arise from an optically thin,
geometrically thick accretion disk, which follows these scalings.  We will
therefore adopt $\phi_{P}=\phi_{\varrho}= \phi_{B}^2= M^{-1}$ as our
fiducial values.  Only if the innermost disk is of the standard gas
pressure dominated \citet{shakura:73} type (which might be the case in low
$\dot{m}$, low $\alpha$ accretion disks in AGNs, \citealt{frank:00}) does
this scaling differ slightly: here, $P\propto B^2 \propto M^{-9/10}$ and
$\varrho\propto M^{-7/10}$.

Jets emit synchrotron radiation from a powerlaw distribution of electrons:
\begin{equation}
	dn/d\gamma = C \gamma^{-p}
\end{equation}
within an energy range $\gamma_{\rm min} < \gamma < \gamma_{\rm max}$.
Here, $\gamma$ is the particle Lorentz factor, $C$ the normalisation
constant, and $p$ the powerlaw index.  The production of powerlaw
distributions is a universal property of diffusive shock acceleration, and
we will assume that the fundamental powerlaw parameters $p$ and
$\gamma_{\rm min}$ are universal in relativistic jets as well.  The
observations from the optically thin part of jet spectra typically give $p
\gtrsim 2$.  In the following, we will take $p=2$ as our fiducial value for
numerical examples.  Since for spectra with $p \geq 2$ the high energy
cutoff is dynamically unimportant, we will not be concerned with its
behavior.  $C$ is then directly proportional to the pressure in
relativistic particles, and we can once again write $C =
\phi_{C}(M)\,\psi_{C}(\chi,\dot{m},a)$.  It is reasonable and customary
to assume that the relativistic powerlaw particle distribution is injected
at some (unknown) fraction of equipartition with the magnetic field
pressure, so $C \propto B^2$.  Thus, for our fiducial values we have
$\phi_{C}=M^{-1}$.

The functions $\psi(\chi)$ can, in principle, take on rather complicated
behavior, depending on the specific jet model.  We will not be concerned
with the detailed nature of $\psi$, so long as they are mathematically well
behaved (see \S\ref{sec:basicequations} for a definition of what this
means).

\subsection{Dependence on $\dot{m}$}
\label{sec:mdot}
While the main aim of this paper is to derive the scaling relation between
jet radio flux and black hole mass, it is interesting to consider the
dependence on other accretion disk parameters, namely $\dot{m}$ (see also
\S\ref{sec:mdot2}).  The dependence of the fundamental disk parameters on
$\dot{m}$ varies more significantly between different accretion models than
the dependence on $M$:

\citep{shakura:73} showed that in radiation pressure supported disks, the
total and magnetic pressure are independent of $\dot{m}$.  Thus, the
magnetic pressure $B^2$ and the relativistic particle pressure $C$ in the
jet are also independent of $\dot{m}$.  The mass density in the disk, on
the other hand, should follow $\varrho\propto \dot{m}^{-2}M^{-1}$, which
might or might not affect the mass loading and thus the Lorentz factor of
the jet.

In mechanically cooled accretion disks (e.g., ADAFs), the pressure $P$ and
particle density $\varrho$ are directly proportional to $\dot{m}$.  In jets
launched from such flows, we thus have $B^2 \propto C \propto
\dot{m}M^{-1}$.

We can derive the same scaling if we simply assume that the mechanical jet
power $W_{\rm jet}$ should be proportional to the disk luminosity $L_{\rm
disk} \propto \dot{M}=M\dot{m}$: Since the jet power at injection is
carried by internal energy, and since we can assume that the magnetic field
and the relativistic particle pressure are related to each other by some
form of dissipation (e.g., reconnection, shocks), both $C$ and $B$ are
related to the jet power by $B^2 \propto C \propto W_{\rm jet}/R^2c \propto
\dot{M}/M^2 \propto \dot{m}M^{-1}$.  As mentioned above, we chose this
parametrisation as our fiducial case.

Finally, in standard gas-pressure dominated disks (however, still dominated
by electron scattering, appropriate for the inner gas-pressure dominated
disks of AGNs and X-ray binaries), the pressure follows $P \propto
\dot{m}^{4/5}\,M^{-9/10}$, and thus, for the jet plasma we have $B^2
\propto C \propto \dot{m}^{2/5}M^{-9/20}$, while the mass density follows
$\varrho \propto \dot{m}^{2/5}M^{-7/10}$.

After this excursion into the dependence of accretion disk and jet
parameters on $\dot{m}$, we now proceed to investigate the radiative
characteristics of self-similar jets.

\section{The non-linear scaling of jet flux with black hole mass and
accretion rate}
\subsection{Synchrotron emission from self similar jets}
\label{sec:emission}
The synchrotron self-absorption coefficient is
\begin{equation}
\alpha_{\nu} = A_{p}\,C\,B^{(p+2)/2}\nu^{-(p+4)/2}
\end{equation}
where $A_{p}$ is a proportionality constant weakly dependent on $p$
\citep{rybicki:79}.

For ease of expression, we will present the following analysis in the case
of a jet viewed from side on, however, extension to the case of arbitrary
viewing angles is straight forward, and the result we derive is fully
general.  In the perpendicular case, the expression for $\tau_{\nu}$ takes
on a particularly simple form:
\begin{eqnarray}
	\tau_{\nu} & = & R_{\rm jet}\,\alpha_{\nu}= R_{\rm
	jet}\,A_{p}\,C\,B^{\left(p+2\right)/2}\nu^{-\left(p+4\right)/2}
	\nonumber\\ & = & A_{p}\,M\, \phi_{C}(M,\dot{m},a)\,
	\left[\phi_{B}(M,\dot{m},a)\right]^{\frac{p+2}{2}}\,
	\nu^{-\frac{p+4}{2}}\, \nonumber \\ & & \ \ \
	\psi_{R}(\chi,\dot{m},a)\, \psi_{C}(\chi,\dot{m},a)\,
	\left[\psi_{B}(\chi,\dot{m},a)\right]^{\frac{p+2}{2}} \nonumber\\ &
	= & \Phi(M,\dot{m},a,\nu)\,\, \Psi(\chi,\dot{m},a)
\label{eq:optical-depth}
\end{eqnarray}
where we define
\begin{eqnarray}
	\lefteqn{\Phi\left(M,\dot{m},a,\nu\right) \equiv M
	\phi_{C}(M,\dot{m},a)
	\left[\phi_{B}(M,\dot{m},a)\right]^{\frac{p+2}{2}}
	\nu^{-\frac{p+4}{2}}}\label{eq:phi} \\
	\lefteqn{\Psi\left(\chi,\dot{m},a\right) \equiv A_{p}
	\psi_{R}(\chi,\dot{m},a) \psi_{C}(\chi,\dot{m},a)
	\left[\psi_{B}(\chi,\dot{m},a)\right]^{\frac{p+2}{2}}}
\end{eqnarray}

The optically thin synchrotron emissivity for a powerlaw distribution of
electrons (well away from the lower and upper cutoff in the energy
distribution) follows
\begin{eqnarray}
	j_{\nu} & = & J_{p}\,C\,B^{\frac{p+1}{2}}\, \nu^{-\frac{p-1}{2}}
	\nonumber \\ & = & J_{p} \phi_{C}(M,\dot{m},a)
	\left[\phi_{B}(M,\dot{m},a)\right]^{\frac{p+1}{2}}
	\nu^{-\frac{p-1}{2}} \nonumber \\ & & \ \ \ \ \ \
	\psi_{C}(\chi,\dot{m},a)
	\left[\psi_{B}(\chi,\dot{m},a)\right]^{\frac{p+1}{2}}
\end{eqnarray}
where $J_{p}$ is a constant weakly dependent on $p$ \citep{rybicki:79}.

For simplicity, we will combine the dependence on the viewing angle
$\vartheta$ due to Doppler beaming and optical depth effects into the
function $\zeta(\vartheta)$.  Because the viewing angle $\vartheta$ and the
Lorentz factor $\Gamma$ are independent of $M$, it follows that
$\zeta(\vartheta)$ must also be independent of $M$, which justifies this
approach in what follows.  The jet surface brightness at a given frequency
$\nu$ is then $S_{\nu}\sim \zeta(\vartheta)\,j_{\nu}\,(1-{\rm
e}^{-\tau_{\nu}})/\alpha_{\nu}$.  The jet flux $F_{\nu}$ is then simply the
surface integral over $S_{\nu}$:
\begin{eqnarray}
	\lefteqn{F_{\nu} = \int_{r_{\rm g}}^{\infty}dr R(r) S_{\nu}(r)
	\approx \zeta(\vartheta)\int_{r_{\rm g}}^{\infty} dr R(r)
	j_{\nu}(r)\frac{1- {\rm
	e}^{-\tau_{\nu}(r)}}{\alpha_{\nu}(r)}}\nonumber\\ & \approx &
	\zeta(\vartheta)\int_{r_{\rm g}}^{\infty} dr
	\,\left[R(r)\right]^2j_{\nu}(r) \frac{1 -
	{\rm e}^{-\tau_{\nu}(r)}}{\tau_{\nu}(r)} \nonumber \\ & \propto &
	\zeta(\vartheta)M^3 \phi_{C}\, \phi_{B}^{\frac{p+1}{2}}
	\nu^{-\frac{p-1}{2}} \int_{1}^{\infty} d\chi \psi_{R}^2 \psi_{C}
	\psi_{B}^{\frac{p+1}{2}} \frac{1 - {\rm e}^{-\Phi\Psi}} {\Phi\Psi}
	\nonumber \\ & \propto & M^3\, \phi_{C}\,\phi_{B}^{\frac{p+1}{2}}\,
	\nu^{-\frac{p-1}{2}}\,
	\Theta\left[\Phi(M,\dot{m},a,\nu),\dot{m},a,\vartheta\right]
\label{eq:peakflux}\end{eqnarray} The integral $\Theta$ depends on $M$,
and $\nu$ only through the combination $\Phi$ from eq.~(\ref{eq:phi}).

\subsection{The relation between $F_{\nu}$ and $M$}
\label{sec:basicequations} 
From eq.~(\ref{eq:peakflux}), we can now work out the non-linear dependence
of $F_{\nu}$ on the central engine mass $M$.  The spectral index
$\alpha\equiv -\partial{\ln{(F_{\nu})}} / \partial{\ln{(\nu)}}$ of the jet
emission is given by
\begin{subequations}
\begin{eqnarray} 
\frac{\partial \ln{(F_{\nu})}}{\partial \ln{(\nu})} & = & - \frac{p-1}{2} +
\frac{\partial \ln{(\Theta)}}{\partial \ln{(\Phi)}} \frac{\partial
\ln{(\Phi)}}{\partial\ln{(\nu)}} \label{eq:nu} \\ & = &
-\frac{p-1}{2}-\frac{\partial \ln{(\Theta)}}{\partial
\ln{(\Phi)}}\left(\frac{p+4}{2}\right) \equiv -\alpha\label{eq:nu1b}
\end{eqnarray}
\end{subequations}
Now taking the partial derivative of eq.~(\ref{eq:peakflux}) with respect
to $M$ and substituting $\partial\ln(\Theta)/\partial\ln(\Phi)$ from
eq.~(\ref{eq:nu1b}), we can write
\begin{subequations}
\begin{eqnarray}
\lefteqn{\frac{\partial \ln{(F_{\nu})}}{\partial \ln{(M)}} = 3 +
	\frac{\partial\ln{\phi_{C}}}{\partial\ln{(M)}} +
	\frac{\partial\ln{\phi_{B}^{\frac{p+1}{2}}}}{\partial\ln{(M)}} +
	\frac{\partial \ln{(\Theta)}}{\partial \ln{(\Phi)}}
	\frac{\partial\ln{(\Phi)}}{\partial\ln{(M)}}} \nonumber \\ & & \ \
	\ \ = \frac{2p+13+2\alpha}{p+4} +
	\frac{\partial\ln{(\phi_{B})}}{\partial\ln{(M)}}
	\left(\frac{2p+3+\alpha p + 2\alpha}{p+4}\right) \nonumber \\ & & \
	\ \ \ \ \ \ \ \, + \frac{\partial\ln{(\phi_{C})}}{\partial\ln{(M)}}
	\left(\frac{5+2\alpha}{p+4}\right) \equiv \xi_{M} \label{eq:mnu1}
\end{eqnarray}
Quite generally, the functions $\phi_{C}$ and $\phi_{B}$ will be simple
powers of $M$ --- for our fiducial assumptions, $\phi_{C}=M^{-1}$ and
$\phi_{B}=M^{-1/2}$, and thus the index $\xi_{M}$ will simply be a
constant:
\begin{eqnarray}
	\xi_{M} & = & \frac{2p+13+2\alpha}{p+4} - \frac{1}{2}
	\left[\frac{2p+3+ (p + 2)\alpha}{p+4}\right] -
	\frac{5+2\alpha}{p+4} \nonumber \\ & \sim &
	\frac{17}{12}-\frac{\alpha}{3} \approx 1.42 - 0.33\alpha
	\label{eq:mnu1b} \hspace{1in}
\end{eqnarray}
\end{subequations}
where the approximate expressions assume $p=2$.  Thus, for any given set of
$\dot{m}$, $a$, and $\vartheta$, $F_{\nu}$ will follow a simple powerlaw
relation in $M$ with powerlaw index $\xi_{M}$
\begin{equation}
	F_{\nu}\propto M^{\xi_{M}} \sim M^{1.42 -
	0.33\alpha}\label{eq:nonlinear}.
\end{equation}
Variations in the other source parameters $\dot{m}$, $a$, the viscosity
parameter $\alpha_{\rm visc}$, and $\vartheta$ will only cause a {\em mass
independent} scatter around this relation.

Remarkably, this result is entirely independent of the functions
$\psi_{f}$.  Given a set of functions $\phi_{f}$, which describe the
dependence of the input conditions in the inner disk on $M$, and given an
observed jet spectrum with spectral index $\alpha$,
eq.~(\ref{eq:nonlinear}) {\em predicts} the scaling of jet flux $F_{\nu}$
with $M$ for any jet model that reproduces this spectral slope.  The only
assumptions that went into the derivation of this result are that {\em a)}
the relevant parameters can be decomposed following
eq.~(\ref{eq:self-similar}), {\em b)} that the high (low) energy cutoffs in
the spectrum are far above (below) the observed spectral band, and {\em c})
that the function $\Theta$ is analytic.  This is what was meant when we
required the functions $\psi_{f}$ to be mathematically well behaved in
\S\ref{sec:selfsimilar}.

Typically, the radio emission from core dominated jets follows a flat
spectrum over many decades in frequency, i.e., $\alpha\sim 0$.  In this
case, it follows for our fiducial parameters that the radio flux $F_{\rm
r}$ depends non-linearly on the mass to the $\xi_{M}=17/12\sim 1.42$ power,
once again, {\em independent} of the jet model, which manifests itself only
through $\psi_{f}$.  \citet{falcke:95} based their adaptation of the
original \citet{blandford:79} model on the assumption that $B\propto
M^{-1/2}$ and $C\propto B^2$.  And indeed, for their specific choice of
$\psi_{f}$, they found $\xi_{M} = 17/12$, which they already showed to be
consistent with observations of flat spectrum radio jets from AGNs and
microquasars \citep{falcke:96}.

As we mentioned before, the fiducial $B^2 \propto C \propto M^{-1}$ scaling
arises in a number of standard scenarios for the inner accretion disk: both
in high efficiency, radiation pressure dominated inner disks and in low
efficiency ADAFs.  The value of $\xi=17/12 - \alpha/3$ is therefore a very
general result, which depends only weakly on the spectral index $\alpha$.
In jets that are launched from standard gas pressure dominated disks that
extend all the way to the innermost stable orbit, the change in the scaling
to $\phi_B^2=\phi_{C}=M^{-9/10}$ leads to a change in the Mass index:
$\xi_{M}=[143+22p-\alpha(14+9p)]/[20(p+4)]\sim 1.56 - 0.23\alpha$, which is
even more non-linear than the standard value of $\xi_{M}\sim 1.42 -
0.33\alpha$ from eq.~(\ref{eq:mnu1b}).  If the magnetic field responsible
for spin extraction from black holes is supported by or anchored in the
inner disk, the same considerations might hold for Blandford-Znajek (1977)
jets.

It is worth noting that this analysis holds even for the case of jets
composed of discrete ejections or internal shocks, if we define $F_{\nu}$
as the time averaged flux or the peak flux.  In fact, because the
derivation of eqs.~(\ref{eq:peakflux}-\ref{eq:mnu1}) did not assume any
specific jet-like geometry, they hold for any synchrotron emitting plasma
with powerlaw spectrum if the source parameters can be described by
eq.~(\ref{eq:self-similar}).

\subsection{The relation between $F_{\nu}$ and $\dot{m}$}
\label{sec:mdot2}
An interesting feature of the derivation of eq.~(\ref{eq:mnu1}) is
that it is modular: any fundamental accretion disk parameter that
enters into the dynamical description of the jet in an invariant
fashion following eq.~(\ref{eq:self-similar}) such that it only
appears in the functions $\phi_f$ and not in $\psi_f$ leads to such a
relation.  For example, if we can separate the dependence of any
dynamical quantity $f$ on the accretion rate $\dot{m}$ into a function
$\phi_{f}$ so that
$f(M,\dot{m},a,r)=\phi(M,\dot{m},a)\,\psi(\chi,a)$, we can derive a
non-linear relation between $F_{\nu}$ and $\dot{m}$ of the
form\footnote{Note that the assumption that the jet Lorentz factor
$\Gamma_{\rm jet}$ is independent of $\dot{m}$, which is implicit in
combining line-of-sight effects into an $\dot{m}$--independent
function $\zeta(\vartheta)$, is not necessarily given.  In the case
where a strong dependence of $\Gamma_{\rm jet}$ on $\dot{m}$ arises,
the complications introduced by Doppler beaming will introduce an
$\dot{m}$-dependent scatter in the $F_{\nu}$--$\dot{m}$ relation,
which could skew the distribution away from the mean scaling index
$\xi_{\dot{m}}$ expected from eq.~(\ref{eq:mdotnu1}).}
\begin{subequations}
\begin{eqnarray}
\lefteqn{\frac{\partial \ln{(F_{\nu})}}{\partial \ln{(\dot{m})}} =
	\frac{\partial\ln{(\phi_{B})}}{\partial\ln{(\dot{m})}}
	\left(\frac{2p+3+\alpha(p + 2)}{p+4}\right)} \nonumber \\ &
	& \ \ \ \ \ \ \ \ \ \ +
	\frac{\partial\ln{(\phi_{C})}}{\partial\ln{(\dot{m})}}
	\left(\frac{5+2\alpha}{p+4}\right) \equiv \xi_{\dot{m}}
	\label{eq:mdotnu1}
\end{eqnarray}
following the same derivation as in eq.~(\ref{eq:mnu1}).  

For our fiducial assumption $\phi_C\propto \phi_B^2 \propto \dot{m}$
(from ADAF type accretion, or the Ansatz $W_{\rm jet}\propto L_{\rm
disk}$) we get
\begin{eqnarray}
	\xi_{\dot{m}} = \frac{2p + (p+6)\alpha + 13}{2(p+4)} \sim
	\frac{17}{12}+\frac{2\alpha}{3} \approx 1.42+0.67\alpha
	\label{eq:mdotnu1b}
\end{eqnarray}
\end{subequations}
where the approximate expressions assume $p=2$. Note that for flat spectrum
sources ($\alpha=0$), the dependence on $\dot{m}$ is the same as that on
$M$, as found by \cite{sams:96}.

Using the accretion disk scaling relations discussed in \S\ref{sec:mdot},
we have presented the powerlaw indices for the jet scaling relations with
mass $M$ and accretion rate $\dot{m}$ expected for these different
accretion modes in Table \ref{tab:scaling}.  The last two columns in this
table show the scaling indices $\xi_{M}$ and $\xi_{\dot{m}}$, such that
\begin{equation}
F_{\nu}\propto M^{\xi_{M}}\dot{m}^{\xi_{\dot{m}}}
\end{equation}

If jet production by black hole spin extraction is invariant under
changes in $a$ in the sense of eq.~(\ref{eq:self-similar}), i.e., if
it depends only trivially on $a$, we can write a relation between
$F_{\nu}$ and $a$ by simply replacing $\dot{m}$ by $a$ in
eq.~(\ref{eq:mdotnu1})

\begin{table}
\begin{center}
\begin{tabular}{l||l|c|c|c}
 & injection mode & $B^2\propto C$ & $\xi_{M}$ & $\xi_{\dot{m}}$ \\
\hline
1 & ADAF & $\frac{\dot{m}}{M}$ & $\frac{17}{12} - \frac{\alpha}{3}$ &
$\frac{17}{12} + \frac{2\alpha}{3}$ \\
2 & rad. press. disk & $M^{-1}$ & $\frac{17}{12} - \frac{\alpha}{3}$ & 0 \\
3 & gas press. disk & $\dot{m}^{\frac{4}{5}}M^{-\frac{9}{10}}$ &
$\frac{187-32\alpha}{120}$ & $\left(\frac{17}{12} +
\frac{2\alpha}{3}\right)\frac{4}{5}$ \\
\hline
4 & $W_{\rm jet}\propto L_{\rm disk}$ & $\frac{\dot{m}}{M}$ & $\frac{17}{12} -
\frac{\alpha}{3}$ & $\frac{17}{12} + \frac{2\alpha}{3}$ \\
\hline
\end{tabular}
\end{center}
\caption{The dependence of $B$ and $C$ on $M$ and $\dot{m}$ and the
scaling indices $\xi_{M}$ and $\xi_{\dot{m}}$ for different accretion
modes (rows 1--3) and for the Ansatz that the mechanical jet
luminosity $W_{\rm jet}$ should be proportional to the disk power
$L_{\rm disk}$ (row 4), assuming $p=2$.\label{tab:scaling}}
\end{table}

\subsection{Optically thin vs.~optically thick emission}
Synchrotron self-absorption is stronger at lower frequencies.  Thus, at
high frequencies the jet must be optically thin even at the location where
it is injected.  Since eqs.~(\ref{eq:mnu1}) and (\ref{eq:mdotnu1}) were
derived without any restrictions on $\tau_{\nu}$, we can use them to infer
the scaling of radiation at high, optically thin frequencies as well, as
long as the other assumptions made above still hold.  The only assumption
which might be violated is that radiative cooling of the electron spectrum
is negligible.  Since modifications by radiative cooling should be visible
as spectral breaks or cutoffs, we will continue to neglect them here,
assuming that a spectral band can be chosen where the spectrum is optically
thin yet unaffected by cooling.

If the jet is injected at a distance $\chi_{\rm i}$ from the black hole
(expressed in dimensionless units), then the frequency $\nu_{\tau=1}$ at
which the jet spectrum becomes optically thin is given by
eq.~(\ref{eq:optical-depth}) by demanding that
$\tau_{\nu}(\chi_{i},M,\dot{m},a)= 1$:
\begin{equation}
	\nu_{\tau}=\left[M\phi_{C}\phi_{B}^{\frac{p+2}{2}}
	\Psi(\chi_{i},...)\right]^{\frac{2}{p+4}}\propto
	\left(M\phi_{C}\phi_{B}^{\frac{p+2}{2}}\right)^{\frac{2}{p+4}}
\end{equation}
Above $\nu_{\tau}$, the spectrum is optically thin and has a spectral index
of $\tilde{\alpha}=\frac{p-1}{2}$ (as can be seen from
eq.~(\ref{eq:peakflux}) by setting $\tau_{\nu} \ll 1$).  Here and below, we
will denote optically thin values by a tilde. E.g., $F_{\tilde{\nu}}$ is
the flux at an optically thin frequency $\tilde{\nu}$.

To derive the scaling indices $\tilde{\xi}_M$ and $\tilde{\xi}_{\dot{m}}$
in the optically thin case, we could go through the same arguments as in
eqs.~(\ref{eq:peakflux}) through (\ref{eq:mdotnu1b}), now imposing that
$\tau_{\tilde{\nu}}=\Phi\Psi \ll 1$.  More easily, however, we can derive
$\tilde{\xi}_{\rm M}$ and $\tilde{\xi}_{\dot{m}}$ by simply replacing
$\alpha$ in eqs.~(\ref{eq:mnu1}) and (\ref{eq:mdotnu1}) by the optically
thin value $\tilde{\alpha}=(p-1)/2$.  This gives
\begin{equation}
	\tilde{\xi}_{M} = 3 + \frac{\partial \ln{(\phi_{C})}}{\partial
	\ln{(M)}} + \frac{p+1}{2}\frac{\partial \ln{(\phi_{B})}}{\partial
	\ln{(M)}} \hspace{5pt}\left[\hspace{3pt} \sim
	\hspace{3pt}\frac{5}{4}\hspace{3pt}\right]
\end{equation}
where the expression in square brackets is valid for $\phi_{C}\propto
\phi_{B}^2\propto M^{-1}$ and $p=2$.

For optically thin jets with $p\sim 2$, \citet{sams:96} suggest that the
observed brightness temperature $\tilde{T}_{\rm b,obs}$ in microquasar and
AGN jets decreases with bolometric luminosity as
$\left(M\dot{m}\right)^{-0.76}$.  The optically thin radio flux then goes
as ${F}_{\tilde{\nu},{\rm obs}} \propto R_{\rm jet}^{2}\tilde{T}_{\rm
b,obs} \propto M^2\,\tilde{T}_{\rm b,obs} \propto M^{1.24}$.  Thus, the
observations give $\tilde{\xi}_{M,{\rm obs}}=1.24$, which coincides nicely
with the theoretical value of $\tilde{\xi}_{M}=1.25$.

For the scaling of optically thin flux $F_{\tilde{\nu}}$ with $\dot{m}$, we
find
\begin{equation}
	\tilde{\xi}_{\dot{m}}= \frac{\partial \ln{(\phi_{C})}}{\partial
	\ln{(\dot{m})}} + \frac{p+1}{2}\frac{\partial
	\ln{(\phi_{B})}}{\partial \ln{(\dot{m})}}
	\hspace{5pt}\left[\hspace{3pt}\sim\hspace{3pt}
	\frac{7}{4}\hspace{3pt}\right]\label{eq:mdotnu2}
\end{equation}
where the expression in square brackets is valid for $\phi_C \propto
\phi_B^2 \propto \dot{m}$ and $p=2$.

For fixed $M$, a change in $\dot{m}$ results in a change to both the
optically thick flux $F_{\nu}$ and the optically thin flux
$F_{\tilde{\nu}}$.  We can thus relate these changes to see how the
optically thick flux varies as a function of the optically thin flux:
\begin{subequations}
\begin{equation} 
	\xi_{\tilde{F}} \equiv \left.\frac{\partial
	\ln{(F_{\nu})}}{\partial \ln{(F_{\tilde{\nu}})}}\right|_{M=const.}
	= \frac{\xi_{\dot{m}}}{\tilde{\xi}_{\dot{m}}}
\end{equation}
where $\xi_{\dot{m}}$ is given in eq.~(\ref{eq:mdotnu1}) and
$\tilde{\xi}_{\dot{m}}$ in eq.~(\ref{eq:mdotnu2}).

If we impose a unique relation between $\phi_C$ and $\phi_B$ (the most
reasonable assumption here is a fixed fraction of equipartition between
relativistic particles and B-field at the base of the jet, which reproduces
our fiducial assumption of $\phi_C \propto \phi_B^2$), we can actually
rewrite $\xi_{\tilde{F}}$ as
\begin{equation}
	\xi_{\tilde{F}} = \frac{1}{p+4}\left[5 + 2\alpha +
	\frac{1-p+2\alpha}{1+p+2\frac{d\ln{\phi_C}}{d\ln{\phi_B}}}\right]
	\label{eq:radioxray2}
\end{equation}
where $\dot{m}$ is now only an implicit parameter.  This implies that the
relation holds for any variation in the jet parameters (whether it is
caused by a change in $\dot{m}$ or any other parameter) as long as it does
not affect the geometry or dimensions of the jet (since both the functions
$\psi_f$ and $R_{\rm jet}$ were kept fixed when deriving
eq.~(\ref{eq:radioxray2})).

For the fiducial assumption of $\phi_C \propto \phi_B^2$,
eq.~(\ref{eq:radioxray2}) reduces to
\begin{equation}
	\xi_{\tilde{F}}=\frac{1}{p+4}\left(4 + 2\alpha +
	\frac{6+2\alpha}{p+5}\right)
	\hspace{5pt}\left[\hspace{3pt}\sim\hspace{3pt} \frac{17 +
	8\alpha}{21}\hspace{3pt}\right] \label{eq:radioxray3}
\end{equation}
\end{subequations}
where the expression in square brackets holds for $p=2$.

Thus, we have a relationship between optically thick and optically thin
flux under the condition of fixed $M$ of the form
\begin{equation}
	F_{\nu}\propto F_{\tilde{\nu}}^{\xi_{\tilde{F}}} \sim
	F_{\tilde{\nu}}^{(17+8\alpha)/21}
\end{equation}
which is remarkably close to the observed correlation between the flat
spectrum radio flux and the X-ray flux observed in the Galactic source GX
339-4 \citep{corbel:00,corbel:03} of $\xi_{\tilde{F},{\rm obs}} \sim 0.7$.

Once again, all of these relations arise independently of the specific jet
model, so long as it produces a spectral index of $\alpha$ in the optically
thick part of the spectrum.  In fact, \cite{markoff:03} find exactly the
same result for $\xi_{\tilde{F}}$ when applying their jet model
\citep{markoff:01} to GX 339-4 as eq.~(\ref{eq:radioxray3}) for the case of
$\alpha=0$ and $p=2$.

\section{Discussion}
\label{sec:discussion}
\subsection{Observational Consequences}
Because the relations of eq.~(\ref{eq:mnu1}) and eq.~(\ref{eq:mdotnu1}) are
model-independent (they only depend on the boundary conditions at the base
of the jet, not on $\psi_{f}$), measurements of $\xi_{M}$, $\xi_{\dot{m}}$,
and $\xi_{\tilde{F}}$ {\em cannot} be used to distinguish between different
jet models.  However, the generality of this result makes such measurements
an even stronger probe of the underlying nature of jet physics:

\begin{itemize}
\item{Observational confirmation of eqs.~(\ref{eq:mnu1}) and
(\ref{eq:mdotnu1}) would prove that jet formation is scale
invariant. On the other hand, if observations can rule out a any
correlation which can be described by these equations, this would
argue strongly against scale invariance.}
\item{Measuring the values of $\xi_{M}$ and $\xi_{\dot{m}}$ would provide
diagnostics of the conditions in the inner disk and at the base of the jet,
i.e., measuring $\xi_{M}$ and $\xi_{\dot{m}}$ could be used to put limits
on $\phi_{f}$.}
\item{Because in low efficiency accretion the accretion rate cannot be
measured directly, it might not be possible to establish a direct
observational correlation between $F_{\nu}$ and $\dot{m}$.  However, if
the above relations hold, any correlation of the optically thick, flat
spectrum jet radio emission with emission at higher frequencies, could be
used to constrain the high energy emission processes (e.g., optically thin
jet emission or bremsstrahlung from an ADAF).}
\item{Measuring the residual spread of $F_{\nu}$ around the predicted
relation could provide a handle on the relative importance of orientation
effects, and thus to measure the mean jet Lorentz factor $\Gamma$.}
\end{itemize}

\subsection{Conclusions}
We have derived the non-linear relation between the observed jet flux at a
given frequency $F_{\nu}$ and the black hole mass $M$.  For scale invariant
jets, the nature of the expression for the jet synchrotron emission makes
it possible to contract all the model dependence into the observable
spectral index $\alpha$.  Thus, for any observed value of $\alpha$, the
derived $F_{\nu}$--$M$ relation is now {\em model independent} --- any jet
model that produces the observed jet spectrum automatically satisfies this
relation.  Given a prescription of the input conditions at the base of the
jet, provided by accretion disk theory, we can thus {\em predict} the
scaling of jet flux with $M$.  Most accretion scenarios produce a scaling
relation of the form $F_{\nu} \propto M^{17/12 - \alpha/3}$.  Thus, for the
optically thick flat spectrum radio emission from core dominate jets we
find $F_{\nu} \sim M^{17/12}$, while for optically thin emission with
$\alpha \sim 0.5$, we find $F_{\nu} \sim M^{5/4}$.  {\em Due to the large
range in black hole mass, this non-linearity makes AGN jets much more radio
loud than microquasar jets.}

This analysis can be extended to any fundamental accretion parameter (e.g.,
accretion rate or black hole spin) if jet dynamics are invariant with
respect to changes in this parameter.  For example, for ADAF-like boundary
conditions at the base of the jet, the scaling with accretion rate
$\dot{m}$ follows $F_{\nu}\sim \left(M\,\dot{m}\right)^{17/12 + 2\alpha/3}$,
and in the flat spectrum case of $\alpha=0$, the dependence on $\dot{m}$
and $M$ is the same: $F_{\nu}\propto \dot{M}^{17/12}$.  Because this result
is model independent, observational measurements of the non-linear scaling
of $F_{\nu}$ with $M$ and $\dot{m}$ are powerful probes of the behavior of
the underlying accretion flows and of the nature of the energy and matter
supply to jets from compact objects.

Clearly, the physics of jet formation is extremely complicated.  For
example, we still do not understand the nature of the radio--loudness
dichotomy in AGNs, and even though they contain black holes of similar
mass, GRS 1915+105 is much more active in the radio band than Cyg X--1.
Nevertheless, independent of the complicated physics of jet formation, the
arguments presented in this paper show that the radio loudness of jets
increases with increasing black hole mass., and thus, that the radio
emission from microquasars should be a much smaller fraction of their
bolometric luminosity than that of radio loud AGNs.
\vspace{0.2in}

\thanks{We would like to thank Eugene Churazov, Tiziana
     DiMattero, Torsten Ensslin, Heino Falcke, Rob Fender, Sera Markoff,
     and Andrea Merloni for helpful discussions and the anonymous referee
     for important suggestions which helped to improve the paper.}


\begin{thebibliography}{}

\bibitem[\protect\citeauthoryear{{Balbus} \& {Hawley}}{{Balbus} \&
  {Hawley}}{1998}]{balbus:98}
{Balbus} S.~A.,  {Hawley} J.~F.,  1998, Reviews of Modern Physics, 70, 1

\bibitem[\protect\citeauthoryear{{Blandford} \& {K\"{o}nigl}}{{Blandford} \&
  {K\"{o}nigl}}{1979}]{blandford:79}
{Blandford} R.~D.,  {K\"{o}nigl} A.,  1979, \apj, 232, 34

\bibitem[\protect\citeauthoryear{{Blandford} \& {Znajek}}{{Blandford} \&
  {Znajek}}{1977}]{blandford:77}
{Blandford} R.~D.,  {Znajek} R.~L.,  1977, \mnras, 179, 433

\bibitem[\protect\citeauthoryear{{Corbel}, {Fender}, {Tzioumis}, {Nowak},
  {McIntyre}, {Durouchoux} \& {Sood}}{{Corbel} et~al.}{2000}]{corbel:00}
{Corbel} S.,  {Fender} R.~P.,  {Tzioumis} A.~K.,  {Nowak} M.,  {McIntyre} V.,
  {Durouchoux} P.,    {Sood} R.,  2000, \aap, 359, 251

\bibitem[\protect\citeauthoryear{{Corbel}, {Nowak}, {Fender}, {Tzioumis} \&
  {Markoff}}{{Corbel} et~al.}{2003}]{corbel:03}
{Corbel} S.,  {Nowak} M.~A.,  {Fender} R.~P.,  {Tzioumis} A.~K.,    {Markoff}
  S.,  2003, \aap, 400, 1007

\bibitem[\protect\citeauthoryear{{Falcke} \& {Biermann}}{{Falcke} \&
  {Biermann}}{1995}]{falcke:95}
{Falcke} H.,  {Biermann} P.~L.,  1995, \aap, 293, 665

\bibitem[\protect\citeauthoryear{{Falcke} \& {Biermann}}{{Falcke} \&
  {Biermann}}{1996}]{falcke:96}
{Falcke} H.,  {Biermann} P.~L.,  1996, \aap, 308, 321

\bibitem[\protect\citeauthoryear{{Fender} \& {Kuulkers}}{{Fender} \&
  {Kuulkers}}{2001}]{fender:01}
{Fender} R.~P.,  {Kuulkers} E.,  2001, \mnras, 324, 923

\bibitem[\protect\citeauthoryear{{Franceschini}, {Vercellone} \&
  {Fabian}}{{Franceschini} et~al.}{1998}]{franceschini:98}
{Franceschini} A.,  {Vercellone} S.,    {Fabian} A.~C.,  1998, \mnras, 297, 817

\bibitem[\protect\citeauthoryear{{Frank}, {King} \& {Raine}}{{Frank}
  et~al.}{2002}]{frank:00}
{Frank} J.,  {King} A.,    {Raine} D.,  2002, {Accretion power in
  astrophysics.~3rd ed.} Cambridge University Press, Cambridge

\bibitem[\protect\citeauthoryear{{Heinz}}{{Heinz}}{2002}]{heinz:02b}
{Heinz} S.,  2002, \aap, 388, L40

\bibitem[\protect\citeauthoryear{{Ho}}{{Ho}}{2002}]{ho:02}
{Ho} L.~C.,  2002, \apj, 564, 120

\bibitem[\protect\citeauthoryear{{Lacy}, {Laurent-Muehleisen}, {Ridgway},
  {Becker} \& {White}}{{Lacy} et~al.}{2001}]{lacy:01}
{Lacy} M.,  {Laurent-Muehleisen} S.~A.,  {Ridgway} S.~E.,  {Becker} R.~H.,
  {White} R.~L.,  2001, \apjl, 551, L17

\bibitem[\protect\citeauthoryear{{Laor}}{{Laor}}{2000}]{laor:00}
{Laor} A.,  2000, \apjl, 543, L111

\bibitem[\protect\citeauthoryear{{Markoff}, {Falcke} \& {Fender}}{{Markoff}
  et~al.}{2001}]{markoff:01}
{Markoff} S.,  {Falcke} H.,    {Fender} R.,  2001, \aap, 372, L25

\bibitem[\protect\citeauthoryear{{Markoff}, {Nowak}, {Corbel}, {Fender} \&
  {Falcke}}{{Markoff} et~al.}{2003}]{markoff:03}
{Markoff} S.,  {Nowak} M.,  {Corbel} S.,  {Fender} R.,    {Falcke} H.,  2003,
  \aap, 397, 645

\bibitem[\protect\citeauthoryear{{McLure} \& {Dunlop}}{{McLure} \&
  {Dunlop}}{2001}]{mclure:01}
{McLure} R.~J.,  {Dunlop} J.~S.,  2001, \mnras, 327, 199

\bibitem[\protect\citeauthoryear{{Narayan} \& {Yi}}{{Narayan} \&
  {Yi}}{1995}]{narayan:95}
{Narayan} R.,  {Yi} I.,  1995, \apj, 444, 231

\bibitem[\protect\citeauthoryear{Rybicki \& Lightman}{Rybicki \&
  Lightman}{1979}]{rybicki:79}
Rybicki G.~B.,  Lightman A.~P.,  1979., {Radiative Processes in
Astrophysics}, Wiley \& Sons, New York

\bibitem[\protect\citeauthoryear{Sams, Eckart \& Sunyaev}{Sams
  et~al.}{1996}]{sams:96}
Sams B.~J.,  Eckart A.,    Sunyaev R.,  1996, \nat, 382, 47

\bibitem[\protect\citeauthoryear{{Shakura} \& {Sunyaev}}{{Shakura} \&
  {Sunyaev}}{1973}]{shakura:73}
{Shakura} N.~I.,  {Sunyaev} R.~A.,  1973, \aap, 24, 337

\bibitem[\protect\citeauthoryear{{Shakura} \& {Sunyaev}}{{Shakura} \&
  {Sunyaev}}{1976}]{shakura:76}
{Shakura} N.~I.,  {Sunyaev} R.~A.,  1976, \mnras, 175, 613

\bibitem[\protect\citeauthoryear{{Woo} \& {Urry}}{{Woo} \&
  {Urry}}{2002}]{woo:02}
{Woo} J.,  {Urry} C.~M.,  2002, \apj, 579, 530

\end{thebibliography}
\end{document}